\documentclass[letterpaper,11pt]{article}
\usepackage{tabularx} 
\usepackage{amsmath}  
\usepackage{graphicx} 
\usepackage[margin=1in,letterpaper]{geometry} 
\usepackage{cite} 
\usepackage[final]{hyperref} 
\hypersetup{
	colorlinks=true,       
	linkcolor=blue,        
	citecolor=blue,        
	filecolor=magenta,     
	urlcolor=blue         
}
\usepackage{blindtext}
\usepackage{amssymb}
\usepackage{amsthm}
\usepackage{authblk}
\usepackage{siunitx}

\begin{document}

\title{\Large Twisted light, a new tool for General Relativity and beyond\footnote{This essay received an Honorable Mention in the 2021 Essay Competition of the Gravity Research Foundation.} \\
\small -- Revealing the properties of rotating black holes with the vorticity of light --}

\author[1]{\normalsize F. Tamburini\footnote{fabrizio.tamburini@gmail.com}}
\author[2]{\normalsize F. Feleppa\footnote{feleppa.fabiano@gmail.com, corresponding author}}
\author[3]{\normalsize B. Thid\'e\footnote{bt@irfu.se}}
\affil[1]{\small ZKM -- Zentrum f\"ur Kunst und Medien, Lorenzstra{\ss}e 19, D-76135 Karlsruhe, Germany}
\affil[2]{\small Institute for Theoretical Physics, Utrecht University, Princetonplein 5, 3584 CC Utrecht, The Netherlands}
\affil[3]{\small Swedish Institute of Space Physics, \si{\angstrom}ngstr\"om Laboratory, Box 537, SE-751 21 Uppsala, Sweden}
\date{\today}

\maketitle

\begin{abstract}
We describe and present the first observational evidence that light propagating near a rotating black hole is twisted in phase and carries orbital angular momentum. The novel use of this physical observable as an additional tool for the previously known techniques of gravitational lensing allows us to directly measure, for the first time, the spin parameter of a black hole. With the additional information encoded in the orbital angular momentum, not only can we reveal the actual rotation of the compact object, but we can also use rotating black holes as probes to test General Relativity.
\end{abstract}

\newpage


Predicted by Einstein's general relativity theory, black holes (BHs) are among the most intriguing objects in our Universe. In the classical general relativistic framework, BHs are characterized solely by their mass, spin and electrostatic charge (which is neutralized to zero in realistic astrophysical settings because of the presence of accreting particles with their own electrostatic charges) \cite{Chandrasekhar}.
As described by Kerr \cite{Kerr}, rotating BHs are massive objects that drag and mix together the surrounding space and time. 

Until now, their experimental confirmation with the usual observational techniques has proved to be extremely difficult.
Over the past decades, researchers tried to develop direct and indirect methods for measuring the spin of astrophysical or galactic BHs like Sagittarius A* \cite{Miller,1996Natur.383..415E,1998ApJ...509..678G}, spanning from X-rays to radio waves, with different techniques that exploit certain properties carried by electromagnetic (EM) waves such as energy, linear momentum and polarization, or by the direct imaging of the supermassive BH in the Messier 87 galaxy within the Virgo cluster \cite{30a,30b,30c,30d,30e,30f}.

It is evident that up to now only a few properties of the EM waves have been exploited in astronomy; all the information carried by EM waves is mainly encoded in the group of standard observables related to the conserved quantities concomitant with the ten-dimensional Poincar\'e group of Noether invariants: energy, the three components of the vector linear momentum, the three components of the pseudovector angular momentum, divided in spin (SAM) and orbital (OAM) \cite{Molina-Terriza&al:NPHY:2007,Torres&Torner:Book:2011}, and the three components of the vector center of energy  \cite{fuschichNikitin2}.

The exploitation of the OAM allows to extract the information encoded in the EM waves by a distant source in a more efficient way not only in telecommunication schemes \cite{Tamburini2012} but also in other fields (for a review see \cite{Torres&Torner:Book:2011,bofabbook}), including observational astronomy \cite{Harwit:APJ:2003,27,Swartzlander:OL:2001,Swartzlander:OL:2005,Foo&al:OL:2005,Lee&al:PRL:2006,Anzolin&al:AA:2008,Tamburini&al:PRL:2006,anzo09} and astrophysics \cite{27a,27b,huptambu,prd17,parruccone}. 

Adopting geometric units $(G=c=1)$, Kerr spacetime, in Boyer-Lindquist coordinates $(t,r,\theta,\phi)$, is described by the following line element \cite{Chandrasekhar}:
\begin{equation}\label{Kerr}
ds^{2}= \frac{\rho}{\Delta}dr^{2} + \rho d\theta^{2} + \frac{\sin^{2}\theta}{\rho^{2}}\left[adt - \left(r^{2}+a^{2}d\phi^{2}\right)\right]^{2} - \frac{\Delta}{\rho^{2}}\left(dt - a\sin^{2}\theta d\phi\right)^{2},
\end{equation}
where the quantities $\rho^{2} \equiv r^{2} + a^{2}\cos^{2}\theta$ and $ \Delta \equiv r^{2} - 2Mr + a^{2}$ depend on the mass $M$ and the rotation parameter $a$ of the spinning BH. 

As it is well-known, the spacetime outside of a spinning BH rotates around the BH itself because of the frame-dragging effect. Because of this, EM waves passing nearby a rotating BH not only experience the gravitational Faraday rotation of the polarization vector that depends on the mass of the BH \cite{faraday} and its rotation parameter $a$, but also acquires OAM independently from the mass of the BH itself thanks to a new general relativistic effect predicted in 2011 and confirmed with numerical simulations \cite{Tamburini2011}.

The Event Horizon Telescope (EHT) team recently released the first picture of the Einstein ring surrounding the shadow of the M$87^{*}$ BH; the image is created by gravitationally lensed beams which, experiencing an anamorphic transformation \cite{Beckwith}, are also accompanied by a polarization rotation \cite{Su} due to the gravitational Faraday effect \cite{Denhen}, image deformation and rotation caused by the lensing effects, as well as other modifications in the phase wavefront, including gravitational Berry phase effects \cite{Carini,Feng,Yang} analogous to that experienced in an inhomogeneous anisotropic medium in a mechanism related to the Pancharatnam-Berry geometric phase \cite{Zela}.

\begin{figure}
\begin{center}
\includegraphics[width=12cm]{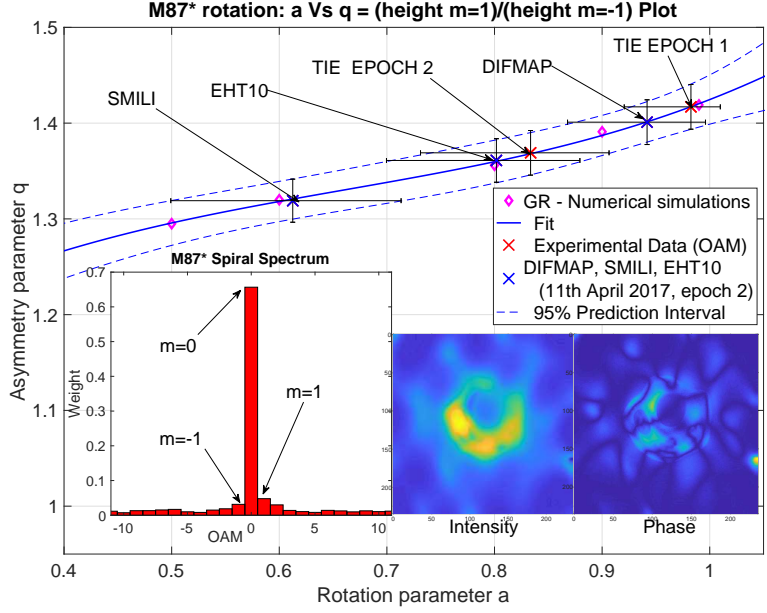}
\end{center}
\caption{Plot of the BH rotation parameter $a$ Vs $q$ for M87*. 
(Main figure): interpolation of the parameter $q$ Vs the rotation parameter $a$  obtained from numerical simulations. Both agree with the experimental data at $95 \%$ confidence level.
(Left, down): the spiral spectrum obtained from the M$87^{*}$ data by reconstructing the spatial phase distribution using well-known techniques based on the Transport of Intensity Equation. (Right, down): intensity and spatial phase distribution of the Einstein ring of M$87^{*}$.}
\label{fig1}
\end{figure}

From the EHT data, the new general relativistic effect hypothesized in \cite{Tamburini2011} was experimentally verified in \cite{Tamburini2020}: light propagating near a rotating BH acquires OAM and reveals directly the BH's rotation. As shown in Fig. 1, light emitted from an accretion disk around a rotating BH is endowed with a specific distribution of OAM that depends on the physical properties of the accretion disk itself and on the rotation of the BH. Spacetime dragging generates a phase-twisting gravitational lens with effects analogous to light traversing a rotating or inhomogeneous anisotropic medium (for further details, see \cite{fermi,gotte,defelice}).
From the analysis of the intensity and phase distribution of the EM field, the OAM can be univocally characterized by the so-called spiral spectrum, the histogram made with the weights of each OAM component present in the received beam (Fig. \ref{fig1}) \cite{OAMspectrum}; it was obtained by decomposing the EM radiation into a set of discrete orthogonal eigenmodes, each carrying a well-defined OAM quantity $m$. Laguerre-Gaussian beams are the most convenient choice and in cylindrical coordinates $(\rho,\phi, z)$ these paraxially approximate modes describe an EM field at distance $z$ with amplitude 
\begin{multline} \label{eq:LG}
 u_{m,p}(r, \theta, z) = \sqrt{\frac{2 p!}{\pi (p + m)!}}
 \frac{1}{w(z)} {\left[\frac{r \sqrt{2}}{w(z)}\right]}^l
 L_p^m \left[\frac{2 r^2}{w^2(z)}\right] 
 \\
 \times \exp {\left[\frac{-\mathrm{i} k r^{2}}{2R(z)}-\frac{r^2}{w(z)^2}\right]} \exp \left[-\mathrm{i} (2 p + m + 1) \arctan
 \left(\frac{z}{z_R}\right)\right] e^{-\mathrm{i} m \varphi },
\end{multline}
where the parameter $m$ represents the azimuthal index of the $z$ component of the OAM, \textit{i.e.}, $m$ represents the number of twists present in the wavefront that has a helical shape in the phase twisted along the $z$ axis, which is the direction joining the BH and the observer. The integer $p$ is the radial node number of the mode, $z_R$ the Rayleigh range of the beam, $R(z)$ the radius of curvature, $w(z)$ the beam waist, $L_p^m$ the associated Laguerre orthogonal polynomial and $\varphi$ is the azimuthal phase of the beam.

The rotation parameter of the BH, $a=0.904 \pm 0.046$ (with $95 \%$ confidence level), is obtained numerically by comparing the ratio $q$ between the $m=1$ and $m=-1$ components from the experimental data of M87* with those obtained from a set of numerical simulations \cite{Tamburini2020}.
If $q>1$, then the rotation is clockwise, whilst $q=1$ indicates no rotation and $q<1$ counterclockwise rotation.
The numerical simulations were used to draw different scenarios seen from an asymptotic observer of an Einstein ring with different emission parameters of the accretion disk, according to EHT results, and varying the inclination of the BH spin $i$ and its rotation parameter $a$.
As output one obtains the field intensity, polarization and phase distributions which allow to determine the content of OAM expected in the observed radiation. Then, from the data provided by the EHT team, the software extracts the radiation intensity and reconstructs numerically the spatial phase distribution of the field in the image plane of an asymptotic observer by using well-known phase reconstruction techniques based on the Transport of Intensity Equation, as explained in the supplementary material of Ref. \cite{Tamburini2020} and references therein. From the analysis of the data it was found that the BH rotates very fast and clockwise with an estimated inclination $i=17^{\circ}\pm 2^{\circ}$.
From these results it is also possible to estimate the amount of energy trapped in the BH rotation, obtaining a value of about $10^{64}$ erg; hence, the M$87^{*}$ rotational energy has the same order of magnitude of the energy emitted by the brightest quasars over Gyr timescales, suggesting that the main engine of quasars and active galactic nuclei depends on the central BH and its rotation.

Despite the success of general relativity in explaining classical gravitational phenomena, several problems at the interface between gravitation and high energy physics still remain open. The novel technique described in this essay can also be used to test gravitational theories beyond general relativity. As it is well known, Heisenberg's uncertainty principle should be modified when gravity is taken into account, leading to a more general formulation also known as the Generalized Uncertainty Principle (GUP). In order to formulate the GUP, a dimensionless parameter $\beta$ is introduced; such a parameter is in principle not fixed by the theory, although it is sometimes assumed to be of the order of unity. This choice, however, renders quantum gravity effects too small to be measured;
therefore, without imposing any condition on the value of the GUP parameter a priori, current experiments can predict larger upper bounds on it, which are compatible with current observations. In order to apply our technique, we first consider the recently proposed GUP-corrected version of a rotating BH, obtained from the line element (\ref{Kerr}) by substituting the BH mare bass $M$ with $M+\frac{\beta}{2M}$ \cite{Carr}; the quantum-corrected solution connects the deformation of the metric of a rotating BH directly to the GUP uncertainty relation. Then, using as probes electromagnetic waves acquiring OAM when lensed by a rotating BH, it is possible to find, from numerical simulations, a relationship between the spectrum of the OAM of light and the
corrections needed to formulate the GUP \cite{FFT}. In this way, we obtain more precise limits on the so-called GUP parameter with respect to those obtained by using BH shadow circularity analyses.

The OAM mechanism described here is in principle valid also for other fields crossing the gravitational field of the Kerr BH. In particular, neutrino fields and gravitational waves may reveal the
hidden latest evolution stages of a supernova during its collapse when the forming Kerr BH is setting its rotation, in an attempt to preserve the total angular momentum while interacting with the surrounding matter and fields; OAM can indeed be induced in both these fields\cite{Barbieri}.

Exploiting OAM in astrophysics allows to reveal the physics of rotating BHs in much more detail than deemed possible before. Our theoretical and observational results open the way to new observational tests of general relativity. Moreover, we have not taken into account the possibility that interactions with plasma turbulence may induce variations in the OAM spectrum of the light at certain frequencies \cite{27a}; although these variations are likely to be much smaller than the gravitational effect, a careful analysis of several different frequency bands could provide very useful information about the turbulence of the disk corona and the traversed interstellar medium.

This new area of research, OAM radio astronomy, could be very promising for the next future.


\end{document}